\documentclass{article}

\usepackage{arxiv}
\usepackage[utf8]{inputenc}
\usepackage[T1]{fontenc}
\usepackage{hyperref}
\usepackage{url}
\usepackage{booktabs}
\usepackage{amsfonts}
\usepackage{amsmath}
\usepackage{nicefrac}
\usepackage{microtype}
\usepackage{cleveref}
\usepackage{graphicx}
\usepackage[numbers]{natbib}
\usepackage{doi}
\usepackage{listings}
\usepackage{xcolor}

\usepackage{array}
\usepackage[caption=false,font=normalsize,labelfont=sf,textfont=sf]{subfig}
\usepackage{textcomp}
\usepackage{stfloats}
\usepackage{verbatim}
\usepackage[linesnumbered,ruled,vlined]{algorithm2e}
\usepackage{tabularx}
\usepackage{adjustbox}
\usepackage{float}
\usepackage{changepage}
\usepackage{multirow}

\title{ReSCom: A Reconfigurable Spiking Neural Network Accelerator Using Stochastic Computing}

\date{}

\author{ Ali Alipour~Fereidani~\textsuperscript{\href{https://orcid.org/0009-0000-6769-0059}{\includegraphics[width=8pt]{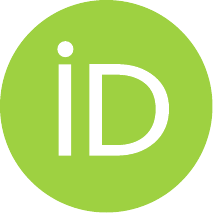}}}\\
	High-Performance Embedded Architecture Laboratory (HiPEAL)\\
	School of Electrical and Computer Engineering\\
	College of Engineering, University of Tehran\\
	Tehran, Iran \\
	\texttt{ali.alipour.f@ut.ac.ir} \\
	\And
	Mohammad~Rasoul Roshanshah~\textsuperscript{\href{https://orcid.org/0000-0001-5650-7925}{\includegraphics[width=8pt]{orcid.pdf}}}\\
	High-Performance Embedded Architecture Laboratory (HiPEAL)\\
	School of Electrical and Computer Engineering\\
	College of Engineering, University of Tehran\\
	Tehran, Iran \\
	\texttt{mrroshanshah@ut.ac.ir} \\
	\And
	Saeed Safari~\textsuperscript{\href{https://orcid.org/0000-0001-6940-591X}{\includegraphics[width=8pt]{orcid.pdf}}}\\
	High-Performance Embedded Architecture Laboratory (HiPEAL)\\
	School of Electrical and Computer Engineering\\
	College of Engineering, University of Tehran\\
	Tehran, Iran \\
	\texttt{saeed@ut.ac.ir} \\
}

\renewcommand{\shorttitle}{ReSCom}

\hypersetup{
	pdftitle={ReSCom}
}

\begin{document}
	\maketitle
	
	\begin{abstract}
		Spiking Neural Networks (SNNs) provide an attractive framework for energy-efficient inference due to their event-driven computation and biologically inspired dynamics. However, efficient hardware realization of SNNs remains challenging because neuronal computations incur significant power and area costs, and uncontrolled approximate arithmetic can destabilize recurrent state updates when precision is not properly managed. To address these challenges, this paper presents ReSCom, a reconfigurable SNN accelerator that leverages stochastic computing to reduce hardware complexity while maintaining stable inference. The proposed architecture employs stochastic arithmetic for multiplication operations in neuron dynamics, while preserving exact fixed-point addition/subtraction operations. This stochastic strategy enables runtime trade-offs between accuracy, latency, and energy consumption. A unified reconfigurable neuron design supports Integrate-and-Fire (IF), Leaky Integrate-and-Fire (LIF), and Synaptic neuron models within a single hardware framework. Experimental results for MNIST inference on a Xilinx Artix-7 FPGA show that ReSCom achieves $92.80\%$ classification accuracy while consuming just $0.05~\mathrm{mJ}$ of operational energy per image at $100~\mathrm{MHz}$, outperforming the energy efficiency of recent state-of-the-art implementations. Furthermore, managing the stochastic bit-stream length allows explicit, dynamic control over accuracy--latency--energy trade-offs to meet target application constraints.
	\end{abstract}
	
	\keywords{Neuromorphic Computing \and Spiking Neural Networks \and Stochastic Computing \and Reconfigurable Accelerator \and FPGA Implementation \and Low-Power Design}

\section{Introduction}
Neural networks have been widely recognized as indispensable tools for solving complex computational problems, ranging from image recognition and natural language processing to autonomous control systems \cite{lecun2015}. While traditional Artificial Neural Networks (ANNs) have achieved remarkable performance, they often require substantial computational resources and high power consumption. To address these inefficiencies, Spiking Neural Networks (SNNs)—often referred to as the third generation of neural networks—have emerged as a promising alternative in neuromorphic computing. Inspired by biological neural processes, SNNs encode and transmit information via discrete spikes rather than continuous values. This event-driven processing mechanism enables sparse computation, offering the potential for significantly higher energy efficiency and lower latency compared to conventional architectures \cite{s41467-1}, \cite{merolla2014}.

However, the physical realization of SNNs on hardware platforms
presents significant challenges.
Despite their theoretical efficiency, practical implementations
can become resource-intensive due to the arithmetic complexity
of neuron dynamics, particularly the multiplication-intensive
state update operations required in models such as LIF and
Synaptic neurons \cite{indiveri2015}.
In large-scale networks, these neuron-level computations may
consume considerable silicon area and power, creating
bottlenecks for deployment on resource-constrained edge
devices such as wearables and mobile platforms \cite{sze2017}.
Consequently, there is a critical need for hardware design
strategies that reduce the cost of neuron dynamics while
maintaining controlled numerical behavior for high-accuracy
inference \cite{hennessy2018}.

To mitigate the hardware overhead of arithmetic operations, Stochastic Computing (SC) has been investigated as a promising alternative to traditional binary logic. SC represents continuous values as streams of random bits, allowing complex arithmetic operations—such as multiplication—to be performed using simple logic gates (e.g., AND gates) rather than area-intensive binary multipliers \cite{alaghi2013}. Several studies have demonstrated that SC can drastically reduce the energy footprint of neural accelerators \cite{ardakani2017}. Nevertheless, adopting SC in SNNs is not without difficulties; the stochastic nature of the computation introduces approximation errors, and long bit-streams can increase latency \cite{alaghi2018}. Prior works have attempted to mitigate these issues, yet finding an optimal balance among hardware efficiency, latency, and classification accuracy remains an open research problem, particularly for complex or adaptable network architectures \cite{kim2016}, \cite{ardakani2019}.

In this paper, we address these challenges by proposing a hardware-efficient SNN architecture that leverages a hybrid computational approach. Unlike fully stochastic designs that may suffer from excessive error, our approach utilizes stochastic logic strictly for neuronal multiplication—the most resource-heavy operation—while maintaining precise arithmetic for synaptic weight accumulation and other addition/subtraction operations. Furthermore, we introduce a reconfigurable neuron model capable of dynamically adapting between Integrate-and-Fire (IF), Leaky Integrate-and-Fire (LIF), and Synaptic behaviors. The stochastic computation flexibility allows our approximate hardware to be tuned for specific application requirements, balancing the trade-off between accuracy and computational simplicity. By integrating a tunable stochastic multiplier with a reconfigurable neuron core, this work paves the way for scalable, low-power neuromorphic systems suitable for FPGA implementation.

The specific contributions of this paper are summarized as follows:
\begin{itemize}
	\item A hybrid stochastic-deterministic computation framework that selectively replaces multiplication operations in neuron dynamics with stochastic units while preserving fixed-point arithmetic for accumulation, reducing DSP utilization and power consumption.
	\item A reconfigurable neuron architecture that unifies IF, LIF, and Synaptic models within a single hardware core, enabling runtime reconfiguration without duplicating hardware resources.
	\item A systematic study of the trade-off between stochastic bit-stream precision and SNN inference accuracy, providing design guidelines for precision tuning in FPGA implementations.
\end{itemize}

The remainder of this paper is organized as follows: Section II reviews related work on spiking neural networks and the principles of stochastic computing. Section III details the proposed methodology, including the reconfigurable neuron architecture and the stochastic multiplier design. Section IV presents the experimental setup, hardware implementation details, and evaluation results. Finally, Section V concludes the paper and outlines directions for future research.

\section{Background and Related Work}

\subsection{Spiking Neural Networks}

The increasing deployment of neural networks in energy-constrained
applications, such as autonomous systems and robotics, has intensified
concerns regarding power consumption.
Spiking Neural Networks (SNNs) have emerged as a promising alternative
to conventional neural networks by adopting event-driven
computation inspired by biological neural systems \cite{2005.01467}.
By operating through sparse spike-based communication, SNNs can significantly reduce 
energy consumption compared to
always-active architectures \cite{s41467-1, s41467-2, s41467-3, s41467-4}.

A defining characteristic of SNNs is their temporal sparsity.
Unlike traditional neural networks, where all neurons continuously
compute, SNNs activate neurons only when spikes occur.
Appropriate tuning of neuron
parameters can ensure that only a subset of neurons is active at a given, 
resulting in improved energy efficiency \cite{s41467}.

\begin{figure}
	\centering
	\includegraphics[width=0.8\columnwidth]{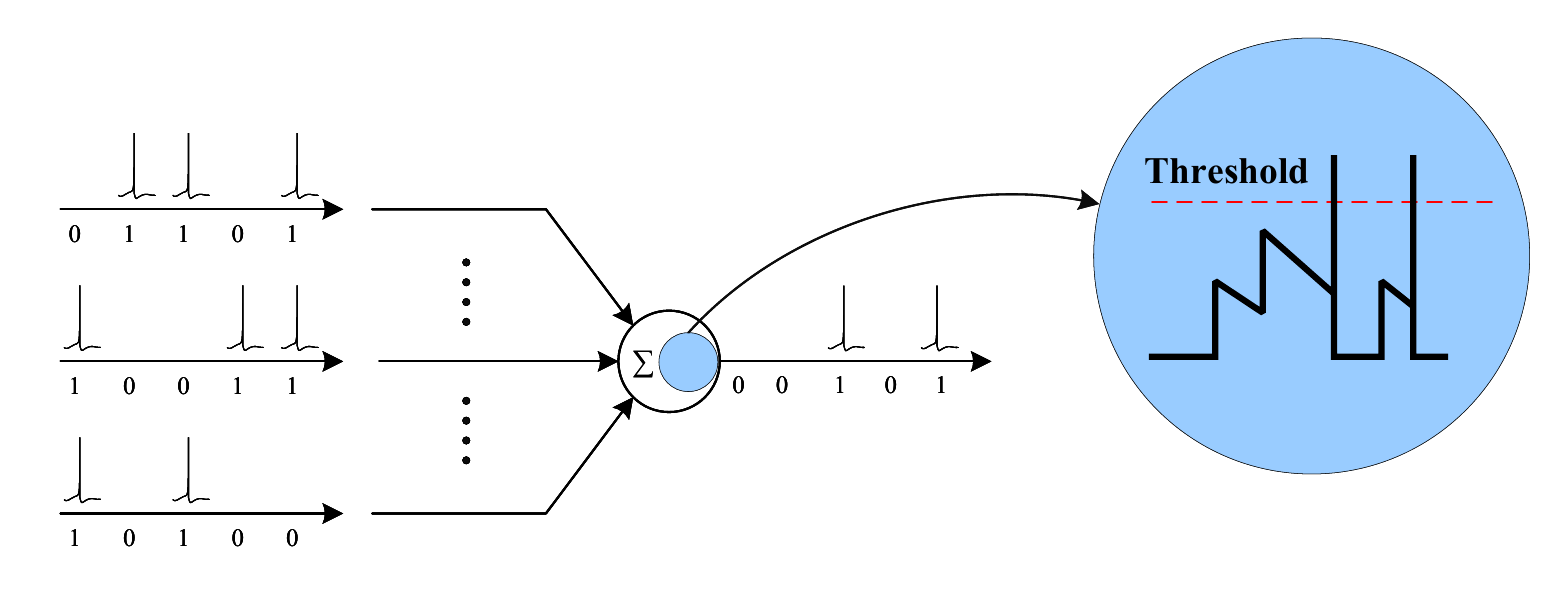}
	\caption{Basic operation of a spiking neuron.}
	\label{fig_1}
\end{figure}

Building upon this temporal sparsity, SNNs encode information using discrete spikes occurring at
specific time instants, as shown in Fig.~\ref{fig_1} \cite{2005.01467}.
Each neuron integrates incoming spikes and emits an output spike when
its membrane potential exceeds a threshold.
This temporal dimension enables SNNs to exploit both spike timing and
spike rate for computation \cite{2005.01476.180}.

\begin{figure}
	\centering
	\includegraphics[width=\columnwidth]{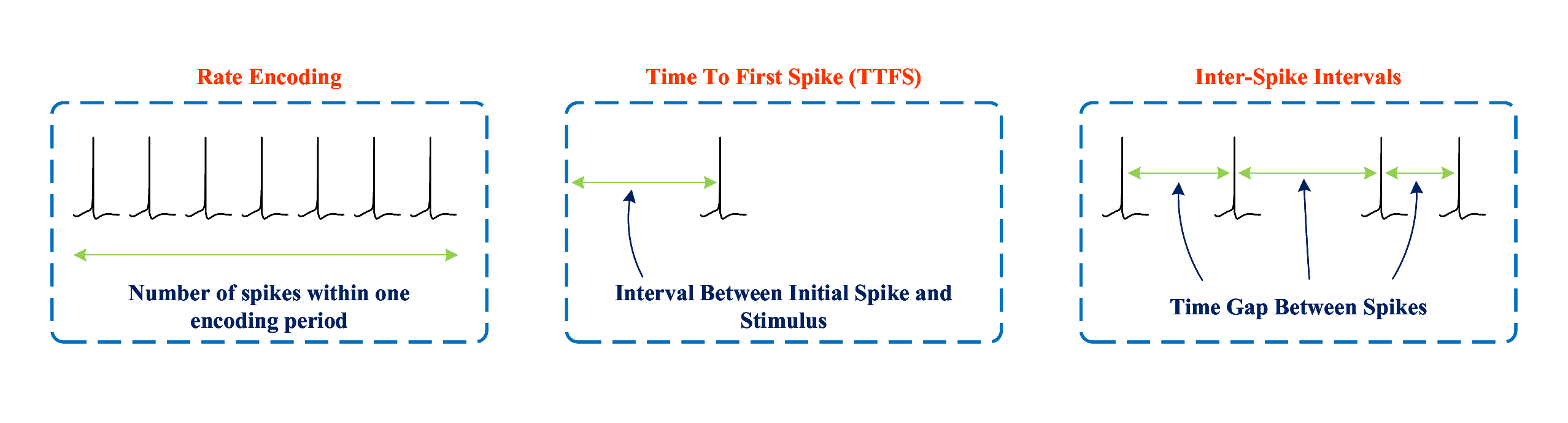}
	\caption{Rate-based and temporal encoding strategies in SNNs.}
	\label{fig_2}
\end{figure}

Several encoding schemes have been proposed to map continuous-valued
inputs to spike trains.
Rate encoding represents information using the number of spikes within
a time window, while temporal encoding methods, such as Time-to-First
Spike (TTFS) and Inter-Spike Interval (ISI), exploit precise spike timing to achieve improved biological plausibility
\cite{Energy_efficient_analog_spiking, 2005.01476.154}.
Fig.~\ref{fig_2} summarizes these encoding strategies
\cite{Guo2021}.

Multiple neuron models have been developed to describe spiking dynamics,
offering different trade-offs between biological realism and
computational complexity \cite{fncom}.
Among these, the Integrate-and-Fire (IF), Leaky Integrate-and-Fire (LIF),
and Synaptic neuron models are widely adopted due to their relative simplicity
and suitability for hardware implementation.
The IF model integrates input current until a threshold is reached
\cite{S00422}, while the LIF model extends this behavior by introducing
membrane leakage \cite{nd_ch1}.
More complex models, such as the Izhikevich neuron, capture a wide range
of biological firing patterns but incur a higher computational cost
\cite{mathmatic}.

The Synaptic neuron model further extends the LIF dynamics by introducing an
explicit synaptic current state, resulting in a second-order
system that more accurately captures synaptic filtering effects.
This model improves temporal representation at the cost of increased
state complexity, making it particularly sensitive to numerical
approximation errors in hardware implementations.

\subsection{Stochastic Computing}

As neural networks scale in size and complexity, parallelization of
arithmetic operations becomes essential but often leads to prohibitive
hardware cost \cite{TNNL-2}.
Conventional fixed-point arithmetic operations such as addition and multiplication 
requires complex and power-hungry circuits
\cite{TNNL-26, TNNL-27, TNNL-28, TNNL-29}.
Stochastic computing has been proposed as an alternative paradigm
that represents values using probabilistic bit-streams rather than
binary words.

\begin{figure}
	\centering
	\includegraphics[width=0.8\columnwidth]{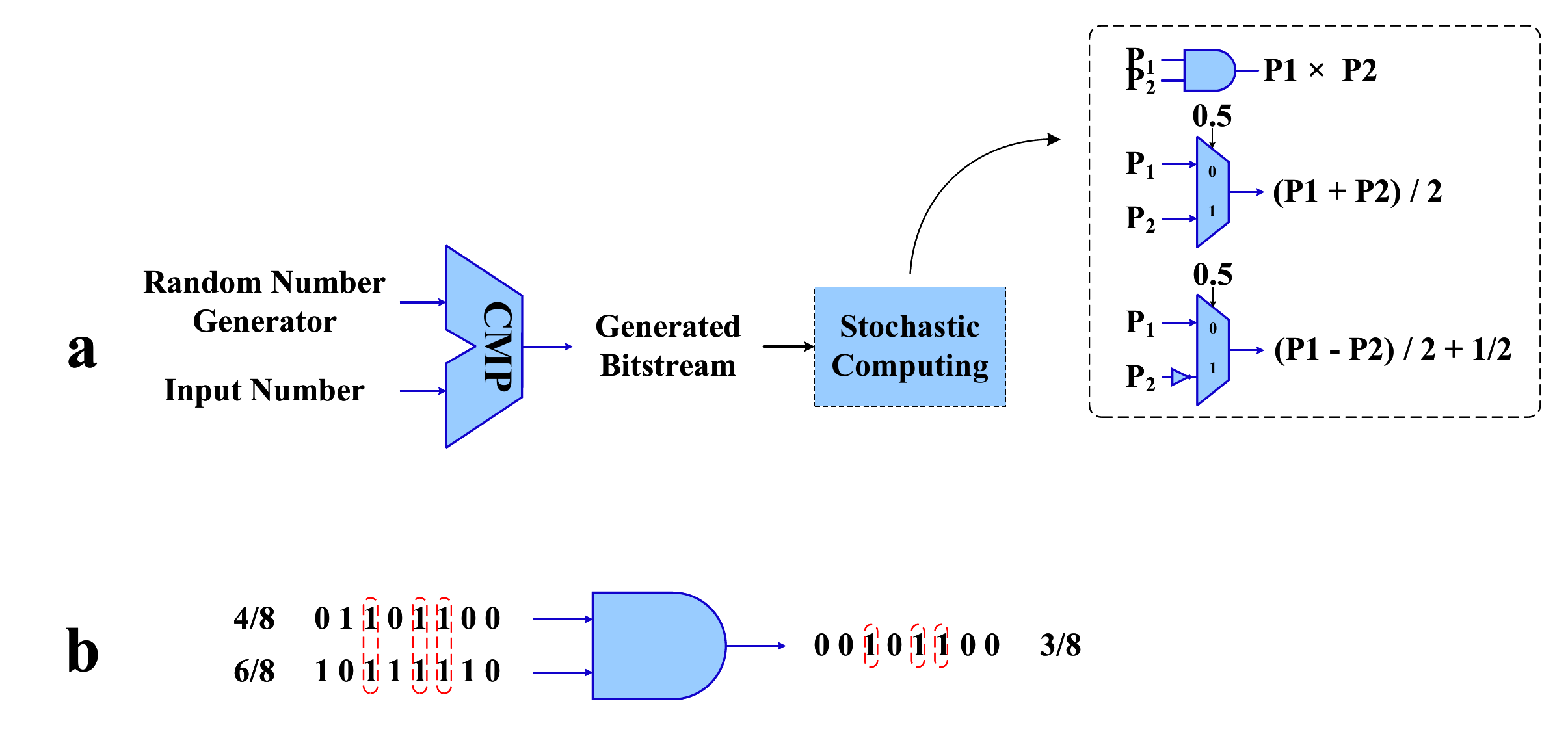}
	\caption{Stochastic computing principles:
		(a) stochastic number representation and arithmetic,
		(b) multiplication using bitwise AND.}
	\label{fig_3}
\end{figure}

In SC, numbers are encoded as the proportion of ones in a stochastic
bit-stream, typically normalized to the range $[0,1]$.
As illustrated in Fig.~\ref{fig_3}(a), arithmetic operations can be realized
using simple logic gates, significantly reducing hardware complexity.
In particular, multiplication can be implemented using a single AND
gate, as illustrated in Fig.~\ref{fig_3}(b), making SC attractive for
energy-efficient hardware implementations \cite{TNNL-31}.

While stochastic computing introduces approximation noise, it offers
inherent robustness against hardware variations and transient faults due to
its probabilistic nature \cite{TNNL-31}.
However, achieving high numerical precision often necessitates long bit-streams, which leads to increased latency.
These limitations motivate hybrid approaches that selectively apply
stochastic computing where its advantages outweigh its drawbacks.

\subsection{Random Number Generation}

Random number generation is a fundamental requirement for stochastic
computing.
Among hardware-based solutions, Linear Feedback Shift Registers (LFSRs)
are widely used due to their simplicity, efficiency, and configurability.
By adjusting the feedback polynomial and seed values, LFSRs can generate
diverse pseudo-random sequences suitable for stochastic encoding \cite{frasser2021, lee2024, lee2025}.

\begin{figure}
	\centering
	\includegraphics[width=\columnwidth]{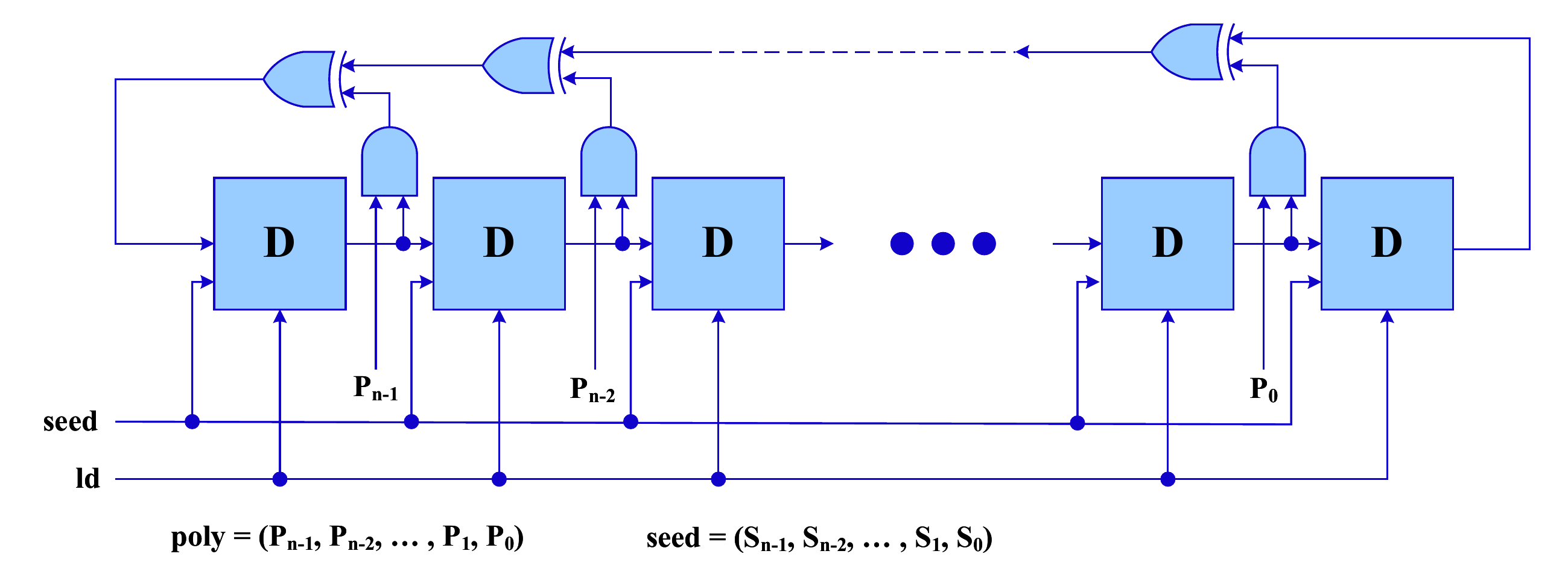}
	\caption{Configurable Linear Feedback Shift Register (LFSR) for
		stochastic number generation.}
	\label{fig_4}
\end{figure}

Fig.~\ref{fig_4} illustrates a configurable LFSR architecture.
The ability to control the statistical properties of generated sequences
makes LFSRs a practical and scalable solution for implementing
stochastic number generators in hardware accelerators.

\section{Proposed Design}
\label{sec:proposed_design}

\subsection{Design Goals and Architectural Rationale}
\label{subsec:design_goals}

The proposed ReSCom architecture is motivated by the growing
demand for energy-efficient and flexible hardware accelerators
for SNNs, particularly in
resource-constrained edge computing platforms.
Although SNNs inherently offer sparse, event-driven computation,
their practical hardware realization remains challenging due
to the high computational cost of neuronal operations and the sensitivity of
neuron dynamics to numerical approximation errors.

Three primary design objectives guide this work.
First, multiplication operations must be implemented in a manner
that significantly reduces power consumption and logic
complexity compared to conventional fixed-point arithmetic.
Second, neuron state updates must remain numerically stable
over time, as spiking neurons rely on recurrent feedback paths
that can amplify approximation errors.
Third, the accelerator should support diverse neuron models rather than 
being restricted to a specific one, thereby accommodating varying requirements for accuracy, 
latency, and biological realism.

To satisfy these objectives, a hybrid arithmetic strategy is
adopted.
Stochastic computing is applied exclusively to neuronal
multiplications, where approximation errors are statistically
bounded and do not accumulate catastrophically.
In contrast, neuron state accumulation and decay operations
are implemented using exact fixed-point arithmetic to preserve
temporal stability.
Furthermore, a reconfigurable neuron architecture is introduced
that supports Integrate-and-Fire, Leaky Integrate-and-Fire, and Synaptic neuron models within a unified hardware
template.
This design enables dynamic tuning of the accuracy--latency--energy trade-off at runtime, rather than fixing the neuron
model during the design phase.

\subsection{Top-Level Architecture}
\label{subsec:system_architecture}

\begin{figure}
	\centering
	\includegraphics[height=0.9\textheight]{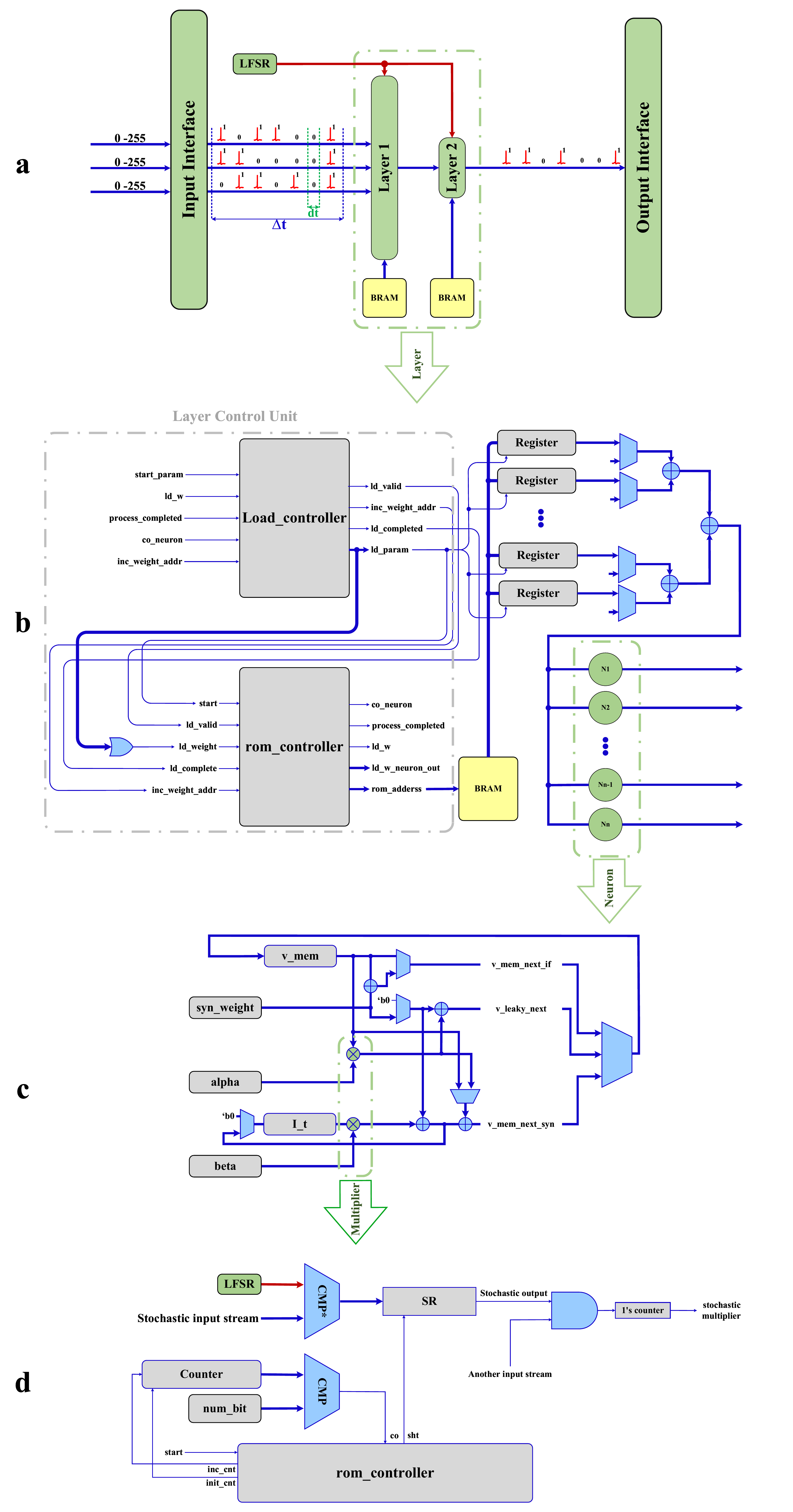}
	\caption{(a) Top-level overview of the proposed ReSCom architecture.
		(b) Layer-level organization of the spiking neural network with a shared
		fully connected (FC) module reused across neurons.
		(c) Reconfigurable neuron micro-architecture supporting IF, LIF, and
		Synaptic neuron models.
		(d) Stochastic multiplier design,
		including bit-stream generation, AND-based multiplication, and
		destochasticization.}
	\label{fig_5}
\end{figure}

Fig.~\ref{fig_5} presents the overall architecture
of the proposed ReSCom design.
The system is organized into three major stages: the
input interface, the network computation unit, and the output
interface.
Together, these components form an end-to-end inference
pipeline that converts static input samples into stochastic
spike streams, processes them using an SNN accelerator, and
generates the final classification results.

\subsubsection{Input Interface and Stochastic Spike Encoding}

As illustrated in Fig.~\ref{fig_5}(a), all input
samples are preprocessed in software prior to hardware
execution.
For example, in the MNIST benchmark, each $28 \times 28$ gray-scale image
is down-sampled to $16 \times 16$ to reduce input dimensionality
and hardware cost.
The resulting 256 pixels are then flattened and normalized to the
range $[0,1]$.

To convert continuous-valued inputs into spike-based
representations suitable for SNN processing, a stochastic
encoding scheme is employed.
Each normalized input value is compared with random numbers
generated from Bernoulli, Normal, or Poisson distributions.
The outcome of each comparison produces a binary spike,
resulting in a stochastic spike stream whose firing rate statistically
approximates the original input magnitude.

\subsubsection{Network Computation}

The network computation unit is implemented on FPGA and
consists of several fully-connected (FC) feed-forward hidden layers and an output layer. 
For the MNIST implementation, the hidden layer contains 256 neurons, while the output layer consists of 10 neurons. 
As shown in Fig.~\ref{fig_5}(b), a shared FC module is reused across neurons within each layer.
This time-multiplexed execution strategy significantly reduces
hardware area and FPGA resource overhead by avoiding replication of synaptic computation logic.

\subsubsection{Output Interface}

Neuron outputs are generated over multiple time steps and
represented as spike trains.
The output interface accumulates spikes from all output neurons
and selects the neuron with the highest spike count as the
predicted class label.
This spike-count-based decoding aligns naturally with rate
encoding schemes and avoids additional post-processing
overhead.

\subsection{Reconfigurable Neuron Micro-Architecture}
\label{subsec:neuron_architecture}

The proposed reconfigurable neuron architecture is illustrated in
Fig.~\ref{fig_5}(c).
The neuron is organized around a shared accumulation datapath,
while model-specific behavior is selected through a 2-bit mode
signal.
This approach enables multiple neuron models to be supported
within a single hardware instance, thereby eliminating the
need for dedicated accelerators for each model. 

Incoming weighted spikes are accumulated using exact fixed-point
adders.
Exact arithmetic is intentionally preserved in the neuron state
update path because spiking neurons rely on recurrent feedback
loops.
Approximation errors introduced in these loops could otherwise
accumulate over time, leading to unstable membrane potentials
and degraded inference accuracy.

When valid synaptic input is available, the neuron updates its
membrane potential and internal state variables.
In the absence of input activity, state registers retain their previous
values, thereby preserving temporal continuity.
This event-driven update mechanism exploits the sparse firing
behavior of SNNs and avoids unnecessary computations.

In IF mode, the neuron integrates weighted synaptic inputs until
the membrane potential exceeds a predefined threshold, at which
point a spike is generated and the membrane potential is reset.
The LIF mode extends this behavior by introducing a leakage
term that causes the membrane potential to decay over time,
thereby improving temporal selectivity.

The Synaptic mode further augments the neuron dynamics by
introducing an additional synaptic current state variable.
This results in a second-order dynamic system that more accurately
captures synaptic filtering effects observed in biological
neurons.
Mode selection is implemented through multiplexer-based control logic,
allowing the neuron to adapt its behavior at runtime according
to application requirements.

\subsection{Stochastic Multiplier Architecture and Precision Trade-Off}
\label{subsec:stochastic_multiplier}

The stochastic multiplier used for synaptic weight computation
is illustrated in Fig.~\ref{fig_5}(d).
Each operand is converted into a stochastic bit-stream using an
LFSR and a comparator.
Multiplication is then performed using a simple bitwise AND
operation. The resulting stochastic bit-stream 
is converted back into a deterministic representation using a one’s counter.

\begin{figure*}
	\centering
	\includegraphics[width=\textwidth]{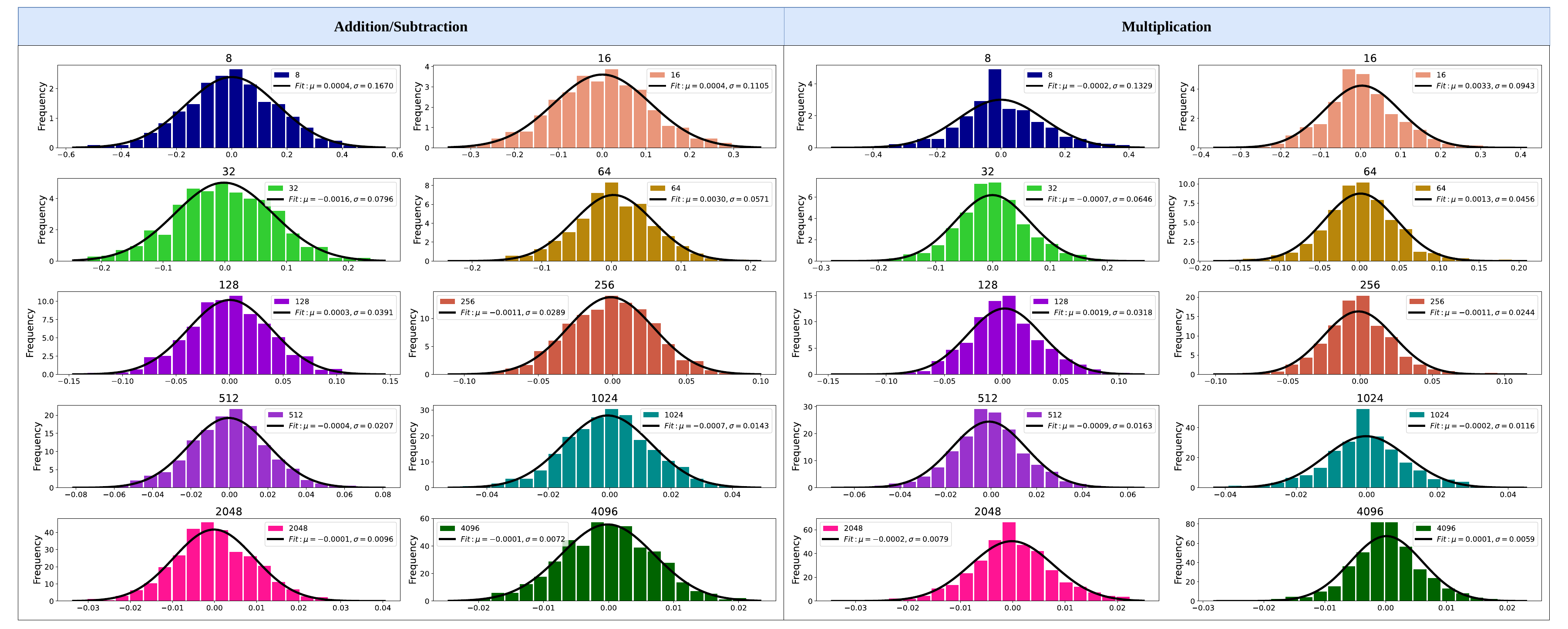}
	\caption{Error distribution of stochastic arithmetic operators for
		different bit-stream lengths.
		Stochastic multiplication exhibits tightly concentrated error
		distributions around zero, while stochastic addition and subtraction
		accumulate significantly larger errors.}
	\label{fig_6}
\end{figure*}

The motivation for restricting stochastic computing to
multiplication is supported by the error distribution analysis
shown in Fig.~\ref{fig_6}.
Stochastic multiplication exhibits tightly bounded error
distributions centered around zero, whereas stochastic addition
and subtraction introduce significantly larger and more
variable errors.
Consequently, stochastic computing is applied only to operations where its
approximation behavior is well controlled.

To recover fixed-point values, the hardware divides the counted
number of ones by the bit-stream length.
For power-of-two bit-stream lengths, this division is
implemented using simple bit shifting, thereby minimizing logic
overhead.
By adjusting the bit-stream length, the proposed architecture
enables explicit tuning of the accuracy--latency--energy trade-off.

\subsection{Memory Organization and Control Logic}
\label{subsec:memory_control}

\begin{figure}
	\centering
	\includegraphics[width=0.6\columnwidth]{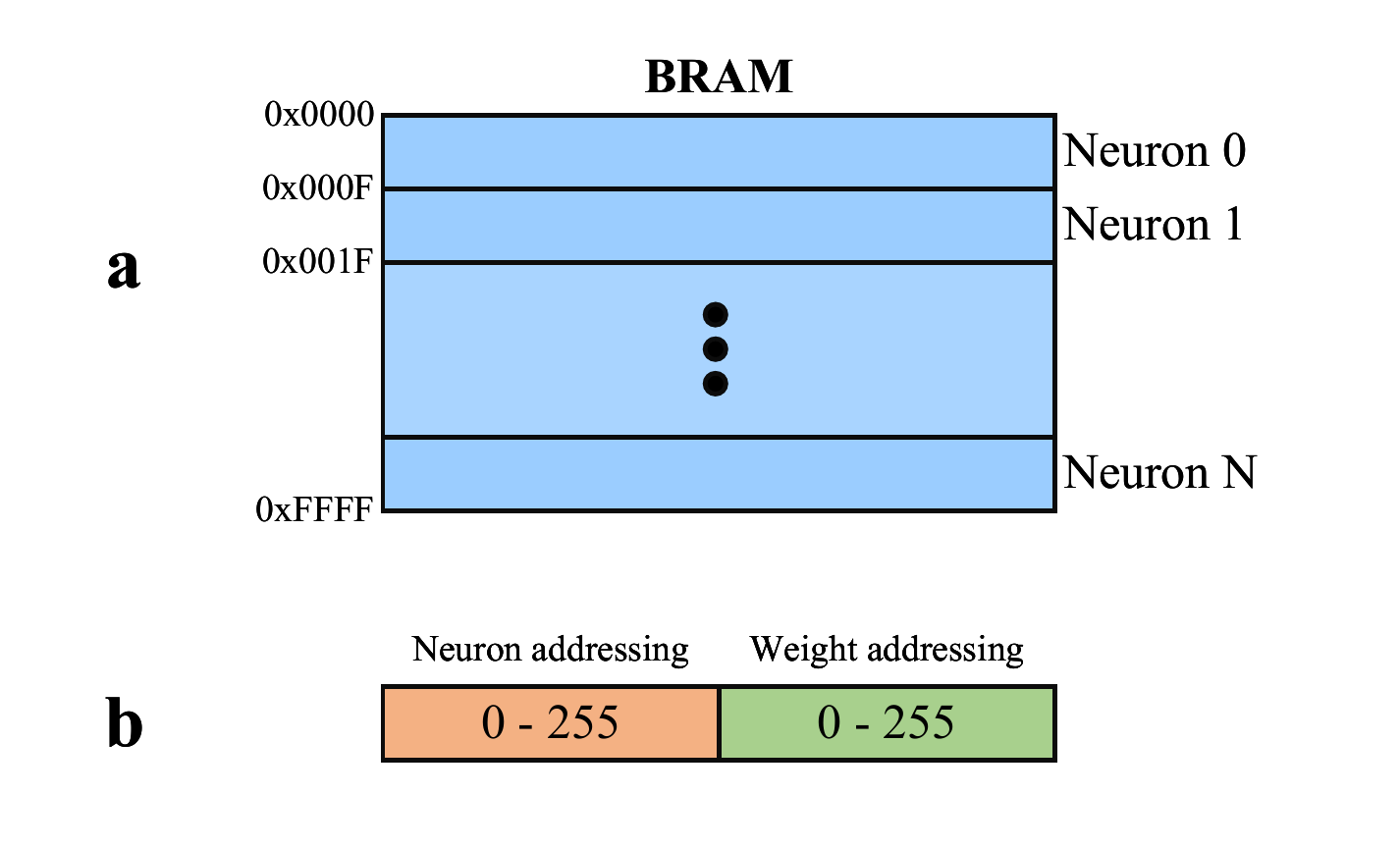}
	\caption{BRAM addressing scheme for synaptic weight storage.
		Neuron indices define address offsets, while contiguous memory
		regions store the corresponding weights.}
	\label{fig_7}
\end{figure}

Synaptic weights are stored in FPGA block RAM (BRAM) and
accessed sequentially during neuron evaluation.
The BRAM addressing scheme is illustrated in
Fig.~\ref{fig_7},
where neuron indices determine address offsets, while contiguous memory
regions store the corresponding synaptic weights.

Two controllers manage memory access and computation.
The load controller is responsible for weight loading and register
initialization, while the ROM controller tracks neuron indices
and corresponding BRAM offsets.
A short delay is inserted between successive neuron evaluations to allow
the stochastic bitstream generation process to complete while keeping the weight registers stable.
This control strategy enables scalable execution of large fully
connected SNN layers while reducing hardware cost.

\section{Results}

This section evaluates the proposed ReSCom accelerator in terms of (i) hardware cost and power consumption, (ii) inference accuracy, and (iii) accuracy--energy trade-offs under different stochastic bit-stream lengths. The primary design objective is to minimize power while avoiding the use of DSP blocks, since DSP-based multipliers typically increase power consumption compared to LUT-based realizations. 

\subsection{Experimental Setup}

MNIST is used as the benchmark dataset \cite{lecun1998mnist}. Each $28\times 28$ image is downsampled to $16\times 16$, flattened into 256 input features, and then encoded into spike trains for network processing. The network consists of one hidden layer with 256 neurons and an output layer with 10 neurons, evaluated over 10 time steps. The training of the spiking neural network is performed using the snnTorch framework \cite{eshraghian2023}. The training and inference configuration used throughout the experiments is summarized in Table~\ref{tab:params}.

\begin{table}
	\centering
	\caption{Network configuration and training parameters used for hardware and software evaluation on the MNIST dataset.}
	\label{tab:params}
	\begin{tabular}{|l|c|}
		\hline
		\textbf{Parameter} & \textbf{Value} \\
		\hline
		Length         & 16 \\
		Input size     & 256 \\
		Hidden size    & 256 \\
		Output size    & 10 \\
		Time steps     & 10 \\
		Batch size     & 128 \\
		Epochs         & 10 \\
		Learning rate  & $1\times10^{-3}$ \\
		Threshold      & $1.0$ \\
		Beta           & $0.98$ \\
		Alpha          & $0.9$ \\
		Neuron mode    & \textbf{\texttt{LIF}} \\
		\hline
	\end{tabular}
\end{table}

\subsection{Multiplier Implementation Impact on the ReSCom Accelerator}

To quantify the system-level hardware impact of stochastic multiplication, the ReSCom accelerator was evaluated across four configurations, varying only the underlying multiplier architecture within the neuron blocks: the proposed stochastic multiplier, a DSP-based multiplier, an array multiplier, and a shift-and-add multiplier. Table~\ref{tab:table_1} summarizes the hardware power consumption, resource utilization, and maximum operating frequency of the entire SNN accelerator under each implementation.

As shown in Table~\ref{tab:table_1}, integrating the stochastic multiplier yields the most energy-efficient configuration, achieving the lowest overall hardware power while maintaining a highly competitive operating frequency. Crucially, the stochastic approach prevents the massive area inflation seen in the array and shift-and-add configurations, which demand significantly higher LUT footprints to implement the neuron updates. While the DSP-based implementation offers a slightly lower LUT count by utilizing dedicated silicon resources, the stochastic design drastically reduces power consumption by over $70\%$ compared to the DSP variant. These results validate the ReSCom architecture's core design choice: using stochastic computing to multiplication operations effectively minimizes total accelerator hardware complexity and power without compromising system frequency.

\begin{table}
    \centering
    \caption{Total hardware resource utilization, power performance, and maximum operating frequency of the ReSCom SNN accelerator on the Artix-7 FPGA across different multiplier configurations.}
    \label{tab:table_1}

        \begin{tabular}{|l|c|c|c|}
        \hline
        \textbf{\shortstack{Accelerator \\ Configuration}} & \textbf{\shortstack{Hardware \\ Power (mW)}} & \textbf{\shortstack{Resource \\ Utilization (LUT)}} & \textbf{\shortstack{Max Frequency \\ (MHz)}} \\
        \hline
        \shortstack{with Stochastic \\ Multiplier} & 7   & 77,848  & 132.42 \\
        \hline
        \shortstack{with DSP \\ Multiplier}        & 24  & 62,549  & 129.35 \\
        \hline
        \shortstack{with Array \\ Multiplier}      & 126 & 411,007 & 57.66  \\
        \hline
        \shortstack{with Shift-Add \\ Multiplier}  & 32  & 305,437 & 110.33 \\
        \hline
        \end{tabular}%
    
\end{table}

\subsection{Inference Accuracy: Hardware vs. Software}

Table~\ref{tab:lif_comparison} reports the inference accuracy on MNIST for the three supported neuron models (IF, LIF, and Synaptic) using a 16-bit stochastic bit-stream length. For each model, both software inference accuracy (floating-point implementation) and measured hardware inference accuracy are reported. The results show that the IF model exhibits the smallest HW--SW accuracy gap, while LIF and Synaptic experience larger degradation under stochastic multiplication. This behavior is consistent with the increased sensitivity of stateful neuron dynamics to approximation noise, particularly when additional internal states are present.

\begin{table}
	\centering
	\caption{Hardware and software inference accuracy on the MNIST dataset for IF, LIF, and Synaptic neuron models using a 16-bit stochastic bit-stream length.}
	\label{tab:lif_comparison}
	\begin{tabular}{|l|c|c|}
		\hline
		\textbf{Model} 
		& \textbf{HW Accuracy (\%)} 
		& \textbf{SW Accuracy (\%)} \\
		\hline
		IF       
		& 96.82 
		& 97.06 					\\
		\hline
		LIF      
		& 92.80 
		& 97.71 					\\
		\hline
		Synaptic 
		& 89.85 
		& 97.49 					\\
		\hline
	\end{tabular}
\end{table}

\subsection{FPGA Implementation and Resource Utilization}

The proposed design is implemented on a Xilinx Artix-7 FPGA. Table~\ref{tab:area} summarizes the required hardware resources. The complete accelerator occupies 77,848 LUTs, corresponding to 57.84\% of the available LUT resources, while consuming 60 BRAM blocks (16.44\%). These results reflect the hardware cost of implementing a fully functional accelerator, including weight storage and the stochastic computing infrastructure.

\begin{table}
	\centering
	\caption{FPGA resource utilization of the proposed ReSCom accelerator implemented on a Xilinx Artix-7 device.}
	\label{tab:area}

		\begin{tabular}{|l|c|c|c|}
			\hline
			\textbf{HW component} & \textbf{LUT} & \textbf{Slice Register} & \textbf{BRAM} 		\\
			\hline
			Single neuron          & 238   & 147   & -   									\\
			\hline
			Hidden layer				   & 71\,570 & 42\,597 & -   								\\
			\hline
			Output layer				   & 3671  & 6172  & -   									\\
			\hline
			LFSR       			   & 35    & 24    & -   									\\
			\hline
			Synaptic weights                & -     & -     & 60  									\\
			\hline
			\shortstack{Complete \\ accelerator}   & \shortstack{77\,848 \\ (57.84\%)} & \shortstack{48\,793 \\ (18.13\%)} & \shortstack{60 \\ (16.44\%)} 	\\
			\hline
			\shortstack{Total FPGA \\ Available}   & 134\,600 & 269\,200 & 365 								\\
			\hline
		\end{tabular}%
	
\end{table}

\subsection{Comparison with SoTA FPGA SNN Accelerators}

Table~\ref{tab:comparison} compares ReSCom with prior FPGA‑based SNN implementations evaluated on the MNIST dataset using the LIF neuron model. For consistency, an operating frequency of 100 MHz is assumed and the overall network energy is estimated using the reported per-image inference latency of 7.24~ms.

Under these conditions, the proposed design consumes approximately $0.05~\mathrm{mJ}$ of energy per image, which is lower than the best previously reported implementation. This highlights a clear latency--energy trade-off: the stochastic multiplier reduces power consumption, but bit-stream generation increases overall execution time. In the targeted operating regime of spiking workloads (where spike rates are typically on the order of a few hertz), this latency increase is not a functional limitation for the intended applications.

\begin{table}
	\centering
	\caption{Comparison of the proposed ReSCom accelerator with state-of-the-art FPGA-based SNN implementations evaluated on the MNIST dataset.}
	\label{tab:comparison}
	\resizebox{\textwidth}{!}{%
		\begin{tabular}{|l|c|c|c|c|c|c|}
			\hline
			\textbf{Design} & \textbf{
				\textbf{\cite{Minitaur_Fpga_Accelerator}}} 				& 
			\textbf{\cite{Energy_efficient_parallel_neuromorphic}} 	& 
			\textbf{\cite{Darwin_neuromorphic_coprocessor}} 		& 
			\textbf{\cite{Fast_energy_efficient_SNN}} 				& 
			\textbf{\cite{Spiker_ISVLSI2022}}  						& 
			\textbf{ReSCom (This work)} 							\\
			\hline
			Clock frequency (MHz) & 75 & 120 & 25 & 100 & 100  & 100 \\
			\hline
			Data format           & 16bit Fixed & 8bit Fixed & 32bit Fixed & 16bit Floating & 16bit Fixed & 16bit Fixed\\
			\hline
			Computing scheme      & Event-Driven & Clock-Driven & Event-Driven & Adaptive Clock/Event-Driven & Clock-Driven & Clock-Driven \\
			\hline
			Neuron model          & LIF & LIF & LIF & LIF & LIF & LIF \\
			\hline
			FPGA platform         & Spartan 6 & Virtex 6 & Spartan 6 & Virtex 7 & Artix 7 & Artix 7 \\
			\hline
			Computation Time	  & 0.40 s/image & 10.08 s/image & 40 ms/image & 3.15 ms/image & 215 $\mu$s/image & 7.24 ms/image \\
			\hline
			Operational Energy	  			  & 0.80 J/image & 1.12 J/image & Not reported & 5.04 mJ/image & 13 mJ/image & 0.05 mJ/image \\
			\hline
			Classification Accuracy              & 92.00\% & 87.7\% & 93.80\% & 92.93\% & 78.58\% & 92.80\% \\
			\hline
		\end{tabular}%
	}
\end{table}

\subsection{Bit-Stream Precision Trade-off}

To characterize the impact of stochastic precision, we evaluate energy consumption and hardware inference accuracy as a function of the stochastic bit-stream length for the IF, LIF, and Synaptic neuron models on the MNIST dataset. As expected, increasing the bit-stream length generally improves accuracy, but increases energy.

\begin{figure}
	\centering
	\includegraphics[width=\columnwidth]{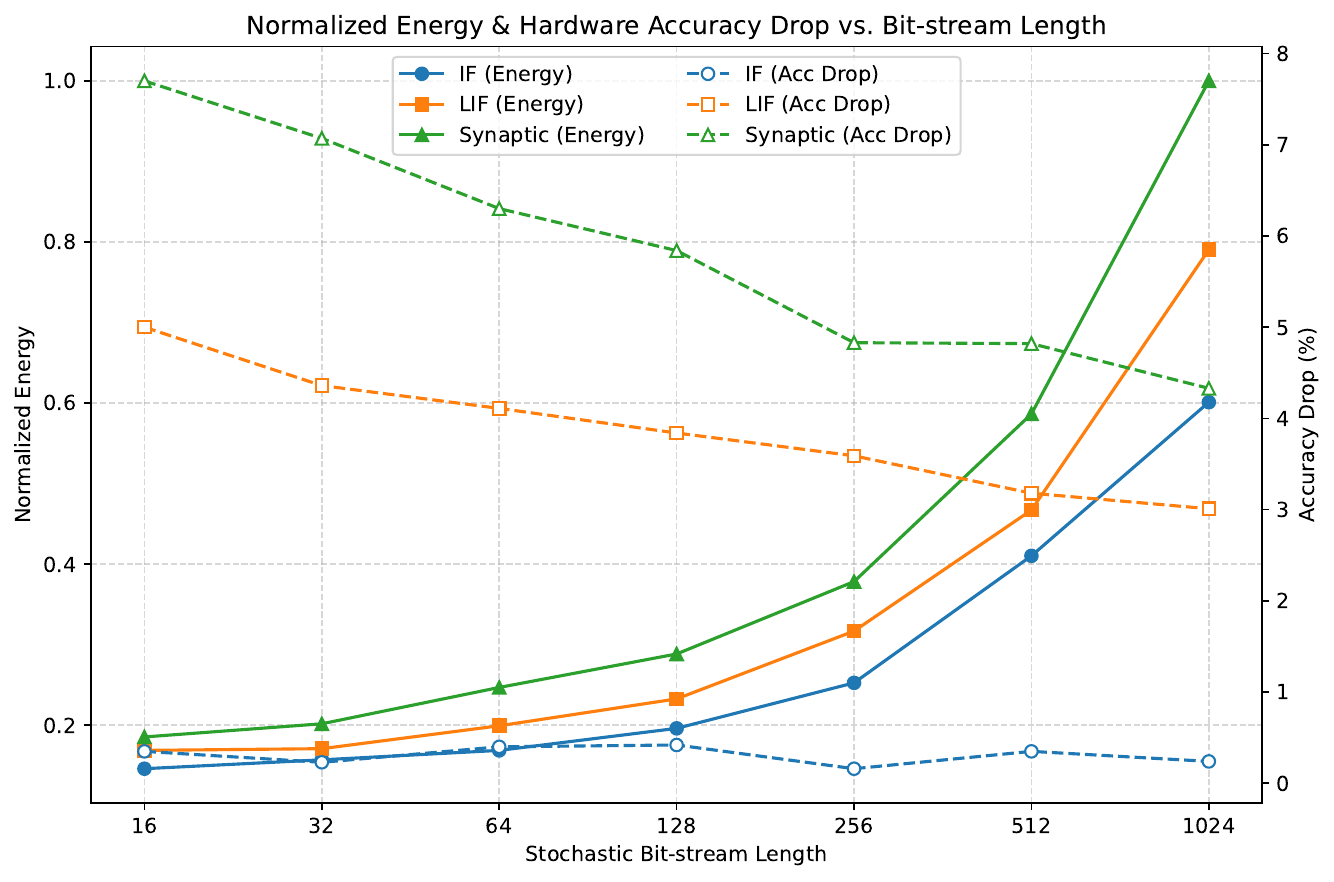}
	\caption{Hardware inference accuracy drop and normalized energy consumption as a function of stochastic bit-stream length for IF, LIF, and Synaptic neuron models on the MNIST dataset. Solid lines denote normalized energy (left axis), while dashed lines represent the accuracy drop compared to the software baseline (right axis).}
	\label{fig_8} 
\end{figure}

Fig.~\ref{fig_8} illustrates the impact of stochastic bit-stream length on hardware inference accuracy drop and energy consumption for different neuron models. 
As the bit-stream length increases, the approximation error introduced by stochastic computing is progressively reduced, leading to improved hardware accuracy and convergence toward the software baseline. 
However, the degree of improvement and the corresponding energy overhead are strongly dependent on the neuron model. 

The IF neuron exhibits minimal sensitivity to stochastic precision, as it involves no internal decay multiplications. This results in a negligible hardware--software accuracy gap, maintaining a minimal accuracy drop across all bit-stream lengths. Correspondingly, its normalized energy consumption remains the lowest among all evaluated models.

The LIF neuron shows moderate sensitivity, since its membrane decay introduces a single stochastic multiplication per time step. For this model, the accuracy drop noticeably improves as the bit-stream length increases, while its energy footprint scales up at a moderate pace.

In contrast, the Synaptic neuron demonstrates the highest sensitivity to stochastic precision. Because it involves two stochastic multiplications and additional state feedback, approximation errors accumulate more rapidly, particularly at shorter bit-stream lengths. Extending the bit-stream effectively recovers the accuracy and reduces the drop, though this comes at the cost of the highest overall energy consumption.

Overall, the results confirm that the proposed reconfigurable architecture enables explicit tuning of the accuracy–energy trade-off by selecting an appropriate stochastic bit-stream length according to the neuron model and application requirements.

\section{Conclusion}

This paper presented ReSCom, a reconfigurable hardware
accelerator for Spiking Neural Networks that combines
stochastic computing with exact fixed-point arithmetic to
achieve energy-efficient and numerically stable inference.
By applying stochastic computation to multiplication-intensive
neuron dynamics while preserving deterministic arithmetic for
state accumulation, the proposed architecture mitigates error
amplification in recurrent feedback paths and significantly
reduces hardware complexity and power consumption.
A unified reconfigurable neuron design supporting IF, LIF, and
Synaptic models enables runtime adaptation of neuron dynamics
to balance biological realism, inference accuracy, and
hardware efficiency.
Implementation on a Xilinx Artix-7 FPGA and evaluation on
benchmark workloads demonstrate competitive classification
accuracy while reducing power consumption compared to DSP-based
SNN implementations.
Future work will investigate extending the SNN architecture to
convolutional and recurrent SNNs, as well as incorporating online learning mechanisms,
such as spike-timing-dependent plasticity, to further enhance adaptability and autonomy.


\begin{thebibliography}{99}
	
	\bibitem{lecun2015}
	Y. LeCun, Y. Bengio, and G. Hinton,
	"Deep learning,"
	\textit{Nature}, vol. 521, no. 7553, pp. 436--444, May 2015.
	
	\bibitem{s41467-1}
	K. Roy, A. Jaiswal, and P. Panda,
	"Towards spike-based machine intelligence with neuromorphic computing,"
	\textit{Nature}, vol. 575, pp. 607--617, Nov. 2019.
	
	\bibitem{merolla2014}
	P. A. Merolla \textit{et al.},
	"A million spiking-neuron integrated circuit with a scalable communication network and interface,"
	\textit{Science}, vol. 345, no. 6197, pp. 668--673, Aug. 2014.
	
	\bibitem{indiveri2015}
	G. Indiveri and S.-C. Liu,
	"Memory and information processing in neuromorphic systems,"
	\textit{Proc. IEEE}, vol. 103, no. 8, pp. 1379--1397, Aug. 2015.
	
	\bibitem{sze2017}
	V. Sze, Y.-H. Chen, T.-J. Yang, and J. S. Emer,
	"Efficient processing of deep neural networks: A tutorial and survey,"
	\textit{Proc. IEEE}, vol. 105, no. 12, pp. 2295--2329, Dec. 2017.
	
	\bibitem{hennessy2018}
	J. L. Hennessy and D. A. Patterson,
	"A new golden age for computer architecture,"
	\textit{Commun. ACM}, vol. 62, no. 2, pp. 48--60, Feb. 2018.
	
	\bibitem{alaghi2013}
	A. Alaghi and J. P. Hayes,
	"Survey of stochastic computing,"
	\textit{ACM Trans. Embed. Comput. Syst.}, vol. 12, no. 2s, Article 92, 2013.
	
	\bibitem{ardakani2017}
	A. Ardakani \textit{et al.},
	"VLSI implementation of deep neural network using integral stochastic computing,"
	\textit{IEEE Trans. VLSI Syst.}, vol. 25, no. 10, pp. 2688--2699, Oct. 2017.
	
	\bibitem{alaghi2018}
	A. Alaghi, W. Qian, and J. P. Hayes,
	"The promise and challenge of stochastic computing,"
	\textit{IEEE Trans. Comput.-Aided Des. Integr. Circuits Syst.},
	vol. 37, no. 8, pp. 1515--1531, Aug. 2018.
	
	\bibitem{kim2016}
	K. Kim, J. Lee, and K. Choi,
	"Dynamic energy-accuracy trade-off using stochastic computing in deep neural networks,"
	in \textit{Proc. 53rd Annu. Design Autom. Conf. (DAC)},
	2016, pp. 1--6.
	
	\bibitem{ardakani2019}
	A. Ardakani \textit{et al.},
	"The synthesis of XNOR recurrent neural networks with stochastic logic,"
	in \textit{Proc. 33rd Conf. Neural Inf. Process. Syst. (NeurIPS)},
	2019, pp. 8442--8452.
		
	\bibitem{Energy_efficient_analog_spiking}
	C. Zhao, J. Li, H. An, and Y. Yi, 
	"Energy Efficient Analog Spiking Temporal Encoder with Verification and Recovery Scheme for Neuromorphic Computing Systems," 
	\textit{Proc. 18th Int'l Symp. Quality Electronic Design (ISQED)}, pp. 138--143, Mar. 2017.
	
	\bibitem{fncom}
	Sanaullah, S., Koravuna, S., Rückert, U., and Jungeblut, T.,
	"Exploring Spiking Neural Networks: A Comprehensive Analysis of Mathematical Models and Applications,"
	\textit{Front. Comput. Neurosci.}, vol. 17, pp. 1215824, Aug. 2023.
	
	\bibitem{S00422}
	A. N. Burkitt,
	"A review of the integrate-and-fire neuron model: I. Homogeneous synaptic input,"
	\textit{Biol. Cybern.}, vol. 95, pp. 1--19, Apr. 2006.
	
	\bibitem{nd_ch1}
	W. Gerstner, W. M. Kistler, R. Naud, and L. Paninski,
	"Neuronal Dynamics: From Single Neurons to Networks and Models of Cognition,"
	\textit{Cambridge University Press}, Chapter 1, Oct. 2013.
	
	\bibitem{mathmatic}
	A. S. Alkabaa, O. Taylan, M. T. Yilmaz, E. Nazemi, and E. M. Kalmoun,
	"An Investigation on Spiking Neural Networks Based on the Izhikevich Neuronal Model: Spiking Processing and Hardware Approach," 
	\textit{Mathematics}, vol. 10, no. 612, pp. 1--16, Feb. 2022.
	
	\bibitem{Guo2021}
	W. Guo, M. E. Fouda, A. M. Eltawil, and K. N. Salama, 
	"Neural Coding in Spiking Neural Networks: A Comparative Study for Robust Neuromorphic Systems," 
	\textit{Front. Neurosci.}, vol. 15, Art. 638474, pp. 1--17, Mar. 2021.
	
	\bibitem{2005.01467}
	M. Bouvier, A. Valentian, T. Mesquida, F. Rummens, M. Reyboz, E. Vianello, and E. Beigne, 
	"Spiking Neural Networks Hardware Implementations and Challenges: A Survey," 
	\textit{J. Emerg. Technol. Comput. Syst.}, vol. 15, no. 2, Art. 22, pp. 1--35, Apr. 2019.
	
	\bibitem{2005.01476.180}
	S. Thorpe, D. Fize, and C. Marlot, 
	"Speed of processing in the human visual system," 
	\textit{Nature}, vol. 381, no. 6582, pp. 520--522, June 1996.
	
	\bibitem{2005.01476.154}
	B. Rueckauer and S.-C. Liu, 
	"Conversion of analog to spiking neural networks using sparse temporal coding," 
	\textit{Proc. Int. Symp. Circuits and Systems (ISCAS)}, pp. 1--5, 2018.
	
	\bibitem{s41467}
	M. Yao \textit{et al.}, 
	"Spike-based dynamic computing with asynchronous sensing-computing neuromorphic chip," 
	\textit{Nat. Commun.}, vol. 15, Art. no. 4464, 2024.
	
	\bibitem{s41467-2}
	C. D. Schuman \textit{et al.}, 
	"Opportunities for neuromorphic computing algorithms and applications," 
	\textit{Nat. Comput. Sci.}, vol. 2, pp. 10--19, Jan. 2022.
	
	\bibitem{s41467-3}
	C. Bartolozzi, G. Indiveri, and E. Donati, 
	"Embodied neuromorphic intelligence," 
	\textit{Nat. Commun.}, vol. 13, Art. no. 1024, Feb. 2022.
	
	\bibitem{s41467-4}
	A. Mehonic and A. J. Kenyon, 
	"Brain-inspired computing needs a master plan," 
	\textit{Nature}, vol. 604, pp. 255--260, Apr. 2022.
	
	\bibitem{TNNL-2}
	A. Krizhevsky, I. Sutskever, and G. E. Hinton, 
	"ImageNet classification with deep convolutional neural networks," 
	in \textit{Proc. Adv. Neural Inf. Process. Syst.}, vol. 25, pp. 1097--1105, 2012.
	
	\bibitem{TNNL-26}
	B. D. Brown and H. C. Card, 
	"Stochastic neural computation. I. Computational elements," 
	\textit{IEEE Trans. Comput.}, vol. 50, no. 9, pp. 891--905, Sep. 2001.
	
	\bibitem{TNNL-27}
	B. D. Brown and H. C. Card, 
	"Stochastic neural computation. II. Soft competitive learning," 
	\textit{IEEE Trans. Comput.}, vol. 50, no. 9, pp. 906--920, Sep. 2001.
	
	\bibitem{TNNL-28}
	J. P. Hayes, 
	"Introduction to stochastic computing and its challenges," 
	in \textit{Proc. Design Autom. Conf. (DAC)}, 2015, p. 59.
	
	\bibitem{TNNL-29}
	A. Alaghi and J. P. Hayes, 
	"Dimension reduction in statistical simulation of digital circuits," 
	in \textit{Proc. Symp. Theory Modeling Simulation, DEVS Integrative M\&S Symp.}, 2015, pp. 1--8.
	
	\bibitem{TNNL-31}
	G. B. Orr and K.-R. Müller, 
	\textit{Neural Networks: Tricks of the Trade}. 
	Berlin, Germany: Springer-Verlag, 2003.
	
	\bibitem{frasser2021}
	C. F. Frasser, M. Roca, and J. L. Rossello,
	"Optimal stochastic computing randomization,"
	\textit{Electronics}, vol. 10, no. 23, p. 2985, 2021.

	\bibitem{lee2024}
	D. Lee, H. Seo, and Y. Kim,
	"Design of an efficient parallel random number generator using a single LFSR for stochastic computing,"
	in \textit{2024 International Conference on Artificial Intelligence in Information and Communication (ICAIIC)}, pp. 775--777, 2024.

	\bibitem{lee2025}
	D. Lee and Y. Kim,
	"Hardware-efficient quantized stochastic computing with reduced precision stochastic number generator and LFSR-based counter,"
	\textit{International Journal of Circuit Theory and Applications}, vol. 53, no. 12, pp. 7281--7294, 2025.

	\bibitem{lecun1998mnist}
	Y. LeCun and C. Cortes, 
	"The MNIST database of handwritten digits," 
	1998. [Online]. Available: http://yann.lecun.com/exdb/mnist/. Accessed: May 2026.
	
	\bibitem{eshraghian2023}
	J. K. Eshraghian \textit{et al.},
	"Training Spiking Neural Networks Using Lessons From Deep Learning,"
	\textit{Proc. IEEE}, vol. 111, no. 9, pp. 1016--1054, Sept. 2023.
	
	
	
	\bibitem{Minitaur_Fpga_Accelerator}
	 D. Neil and S.-C. Liu, 
	 "Minitaur, an Event-Driven FPGA-Based Spiking Network Accelerator," 
	 \textit{IEEE Trans. VLSI Syst.}, vol. 22, no. 12, pp. 2621--2628, Dec. 2014.
	 
	 \bibitem{Energy_efficient_parallel_neuromorphic}
	 Q. Wang \textit{et al.},
	 "Energy Efficient Parallel Neuromorphic Architectures with Approximate Arithmetic on FPGA,"
	 \textit{Neurocomputing}, vol. 221, pp. 146--158, Jan. 2017.
	 
	 \bibitem{Darwin_neuromorphic_coprocessor}
	 D. Ma \textit{et al.},
	 "Darwin: A Neuromorphic Hardware Co-Processor Based on Spiking Neural Networks,"
	 \textit{J. Syst. Archit.}, vol. 77, pp. 43--51, Jan. 2017.
	 
	 \bibitem{Fast_energy_efficient_SNN}
	 S. Li \textit{et al.},
	 "A Fast and Energy-Efficient SNN Processor With Adaptive Clock/Event-Driven Computation Scheme and Online Learning,"
	 \textit{IEEE Trans. Circuits Syst. I, Reg. Papers}, vol. 68, no. 4, pp. 1543--1552, Apr. 2021.
	 
	 \bibitem{Spiker_ISVLSI2022}
	 A. Carpegna, A. Savino, and S. Di Carlo,
	 "Spiker: an FPGA-Optimized Hardware Accelerator for Spiking Neural Networks,"
	 in \textit{Proc. IEEE Comput. Soc. Annu. Symp. VLSI (ISVLSI)}, 2022, pp. 14--19.
	

\end{thebibliography}
\end{document}